\documentclass{article}

\usepackage{epic}
\usepackage{eepic}

\newcommand{\operatorname}[1]{\mbox{#1}}

\mathchardef\ordinarycolon\mathcode`\:     
\mathcode`\:=\string"8000
\def\vcentcolon{\mathrel{\mathop\ordinarycolon}}
\begingroup \catcode`\:=\active
  \lowercase{\endgroup
  \let :\vcentcolon
  }

\newcommand{\nc}{\newcommand}

\nc{\rnc}{\renewcommand}
\nc{\B}{\cal{B}}
\nc{\lbar}[1]{\overline{#1}}
\nc{\bra}[1]{\langle#1|}
\nc{\ket}[1]{|#1\rangle}
\nc{\ketbra}[2]{|#1\rangle\!\langle#2|}
\nc{\braket}[2]{\langle#1|#2\rangle}
\nc{\proj}[1]{| #1\rangle\!\langle #1 |}
\nc{\inner}[2]{(#1,#2)}
\nc{\IP}{\operatorname{IP}}
\rnc{\max}{\operatorname{max}}
\nc{\Prob}{\operatorname{Prob}}
\nc{\Rank}{\operatorname{Rank}}
\nc{\Span}{\operatorname{Span}}
\rnc{\S}{\operatorname{S}}
\nc{\diag}{\operatorname{diag}}
\nc{\sign}[1]{\mbox{sign$(#1)$}}
\nc{\smfrac}[2]{\mbox{$\frac{#1}{#2}$}}
\nc{\Spec}{\operatorname{Spec}}
\nc{\Tr}{\operatorname{Tr}}
\nc{\xor}{\oplus}
\nc{\ox}{\otimes}
\nc{\dg}{\dagger}
\nc{\dn}{\downarrow}
\nc{\cA}{{\cal A}}
\nc{\cB}{{\cal B}}
\nc{\cC}{{\cal C}}
\nc{\cD}{{\cal D}}
\nc{\cE}{{\cal E}}
\nc{\cF}{{\cal F}}
\nc{\cG}{{\cal G}}
\nc{\cH}{{\cal H}}
\nc{\cI}{{\cal I}}
\nc{\cJ}{{\cal J}}
\nc{\cK}{{\cal K}}
\nc{\cL}{{\cal L}}
\nc{\cR}{{\cal R}}
\nc{\cS}{{\cal S}}
\nc{\cX}{{\cal X}}
\nc{\supp}{{\operatorname{supp}}}
\nc{\stab}{{\operatorname{stab}}}
\nc{\lrar}{\longrightarrow}

\def\a{\alpha}
\def\b{\beta}

\def\d{\delta}
\def\e{\varepsilon}

\def\m{\mu}

\def\s{\sigma}

\def\ph{\varphi}

\def\ps{\psi}
\def\o{\omega}

\def\S{\Sigma}

\def\Ph{\Phi}

\nc{\RR}{{{\mathbb R}}}
\nc{\CC}{{{\mathbb C}}}
\nc{\FF}{{{\mathbb F}}}
\nc{\NN}{{{\mathbb N}}}
\nc{\ZZ}{{{\mathbb Z}}}
\nc{\PP}{{{\mathbb P}}}
\nc{\QQ}{{{\mathbb Q}}}
\nc{\UU}{{{\mathbb U}}}
\nc{\Q}{{{\bar{Q}}}}
\nc{\nill}[1]{}

\title{\textbf{Embezzling Entangled Quantum States}}
\author{Wim van Dam\thanks{Computer Science Division, Soda Hall, 
University of California, Berkeley, CA 94720 (USA). Also at MSRI Berkeley and 
HP Palo Alto.
Email: vandam@cs.berkeley.edu.} \and 
Patrick Hayden\thanks{Institute for Quantum Information, Caltech 107-81, 
Pasadena, CA 91125 (USA). Email: patrick@cs.caltech.edu.}}
\date{January 10, 2002}

\begin{document}

\maketitle

\begin{abstract}
We show that in the presence of arbitrary catalysts, any pure bipartite 
entangled state can be converted into any other to unlimited accuracy
without the use of any communication, quantum or classical.
\end{abstract}


\vspace{0.5cm}
\noindent
The interconvertibility of entangled quantum states is an important question 
in quantum information theory, both for its own sake and because of its
connections to quantum error correction \cite{BennettDSW96}, quantum 
cryptography \cite{Ekert91} and quantum communication 
complexity \cite{Ambainis98}.  In 1999, Nielsen and Hardy 
supplied a powerful tool for studying this
problem, in the form of a simple characterization of the 
bipartite pure states convertible into each other using only local operations and 
classical communication (LOCC) \cite{Hardy99,Nielsen99}.  
Building on that work, complete characterizations of the
corresponding probabilistic \cite{Vidal99} and approximate \cite{VidalJN00} 
conversion problems soon followed.
In addition, Jonathan and Plenio discovered the existence of \emph{catalysts:}
states that are recovered once a transformation is complete but whose 
presence allows successful LOCC protocols 
that would not otherwise have been possible \cite{JonathanP99}.

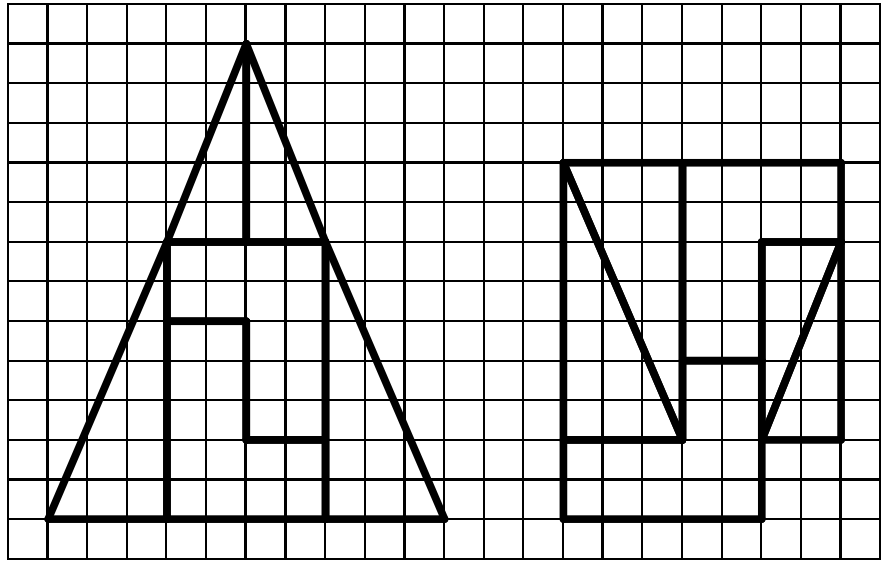
\begin{figure} 

\unitlength = 1.5pt
\begin{center}
\begin{picture}(220,130)(-10,0)
\put(-8,-10){\grid(220,140)(10,10)}
\allinethickness{3pt}
\drawline(100,0)(70,0)(70,70)(100,0)
\drawline(0,0)(30,0)(30,70)(0,0)
\drawline(30,70)(50,70)(50,120)(30,70)
\drawline(70,70)(50,70)(50,120)(70,70)
\drawline(30,0)(70,0)(70,20)(50,20)(50,50)(30,50)(30,0)
\drawline(30,50)(50,50)(50,20)(70,20)(70,70)(30,70)(30,50)

\drawline(130,0)(180,0)(180,40)(160,40)(160,20)(130,20)(130,0)
\drawline(130,20)(130,90)(160,20)(130,20)
\drawline(130,90)(160,90)(160,20)(130,90)
\drawline(160,40)(180,40)(180,70)(200,70)(200,90)(160,90)(160,40)
\drawline(180,20)(200,20)(200,70)(180,20)
\drawline(180,20)(200,70)(180,70)(180,20)
\end{picture}
\end{center}
\caption{An illustration of the `embezzlement effect'.
By a well-chosen rearrangement
we can create the suggestion that the six pieces of the
rightmost figure, with area size $59$, can also be used 
the cover the triangle on the left with its surface of $60$ units.
A similar phenomenon is described in this Letter for the entanglement 
of a distributed quantum state. 
It is shown how we can reorder the amplitudes of an embezzling state 
$\mu$ such that we get a very close approximation of an enlarged state
$\mu\otimes \varphi$, which appears to have significantly more entanglement 
than the original $\mu$. \label{figure}}
\end{figure}

In this Letter, we exhibit a family of bipartite catalysts 
$\{\ket{\m(n)}\}_{n=1}^\infty$ 
such that, for any $\e > 0$ and any bipartite state $\ket{\ph_{AB}}$, the
transformation
\begin{eqnarray}
\ket{\m(n)} &\mapsto& \ket{\m(n)}\ox\ket{\ph_{AB}}
\end{eqnarray}
can be accomplished with fidelity better than $1-\e$ for all sufficiently 
large $n$ without any communication, quantum or classical.
In other words, it is possible to \emph{embezzle} a copy of $\ket{\ph_{AB}}$
from $\ket{\m(n)}$, thereby removing a small amount of entanglement from the
original state, while causing only an arbitrarily small disturbance $\e$ 
to it.  
This embezzlement protocol only requires the two parties $A$ and $B$
to rearrange the coefficients of the $\m(n)$ state such that it 
resembles the desired $\m(n)\otimes \ph_{AB}$.
(An analogy to this phenomenon is illustrated in Figure~\ref{figure}.)
Because the set of states $\{{\m(n)}\}$ can be used to embezzle any target
state $\ph$ to within an arbitrarily high fidelity $1-\e$ that depends
only on the Schmidt rank of $\ph$ and the size $n$, we call the set a
\emph{universal embezzling family}.
It follows trivially that this family can also be used as a catalyst 
to `convert' a now superfluous $\psi_{AB}$ to $\ph_{AB}$ with 
arbitrarily small error.

The index $n$ indicates the Schmidt rank of the specific $\ket{\mu(n)}$, 
and for each $n$ the embezzling state is defined by
\begin{eqnarray}
\ket{\m(n)} &:=& \frac{1}{\sqrt{C(n)}} \sum_{j=1}^n
\frac{1}{\sqrt{j}} \ket{j}_A \ket{j}_B,
\end{eqnarray}
where $C(n) := \sum_{j=1}^n \frac{1}{j}$ is chosen so that
$\ket{\m(n)}$ is normalized.  Now suppose we would like to embezzle
the state $\ket{\ph_{AB}} := \sum_{i=1}^m \a_i \ket{i}_A\ket{i}_B$ from
$\ket{\m(n)}$, where $\ket{\ph_{AB}}$ is written according to its Schmidt decomposition
such that all $\alpha_i$ amplitudes are positive reals.
 This problem is equivalent to creating the state 
$\ket{\o(n)} = \sum_{j=1}^{mn} \o_j \ket{j}_A\ket{j}_B$, which is defined
as the state with the same Schmidt basis and coefficients as 
$\ket{\m(n)}\ox\ket{\ph_{AB}}$ but with the coefficients $\o_j$ in 
decreasing order.
Thus $\ket{\o(n)}$ can be converted into $\ket{\m(n)}\ox\ket{\ph_{AB}}$ 
exactly by local unitary operations alone.  The embezzlement protocol 
will simply consist
of performing these local unitaries because we will show that 
$|\braket{\m(n)}{\o(n)}|$ goes to $1$ as $n$ goes to infinity.

The first step in the proof will be to show that the first $n$ Schmidt 
coefficients of $\ket{\o(n)}$ are smaller than the corresponding ones 
of $\ket{\m(n)}$.  To see this, observe that the first $n$ Schmidt coefficients of 
$\ket{\o(n)}$ are all of the form $\a_i/\sqrt{j C(n)}$, 
where $1 \leq i \leq m$ and $1 \leq j \leq n$.  For a fixed $t$ and $i$, 
we let $N_i^t$ be the number of such coefficients $\a_i/\sqrt{j C(n)}$ 
that are strictly greater than $1/\sqrt{tC(n)}$.
By the restriction $1\leq j < \a_i^2t$, it follows that 
 $N_i^t < \a_i^2 t$ and, since $\sum_{i=1}^m \a_i^2 = 1$, 
we can conclude that $\sum_{i=1}^m N_i^t < t$.
This upper bound on the number of $\o_j$ coefficients that are 
strictly bigger than $1/\sqrt{t C(n)}$ combined with the ordering 
$\o_1\geq\o_2\geq\dots\geq\o_{mn}$ proves that $\o_j \leq 1/\sqrt{jC(n)}$ 
for all $1 \leq j \leq n$.
Consequently, the fidelity between $\ket{\mu(n)}$ and $\ket{\o(n)}$
can be bounded from below by
\begin{eqnarray}
|\braket{\m(n)}{\o(n)}|
&=& \sum_{j=1}^n \frac{\o_j}{\sqrt{jC(n)}}
~ ~\geq ~ ~\sum_{j=1}^n \o_j^2.
\end{eqnarray}
Our next task is to show that this sum is close to $1$ for large $n$. 
Let $\ket{\ps(n)} := \ket{\m(n)}\ox\ket{\Ph^m}$, where 
$\ket{\Ph^m} := \smfrac{1}{\sqrt{m}} \sum_{i=1}^m \ket{i}_A\ket{i}_B$ is the
maximally entangled state of rank $m$.  Then $\o(d)_A \succ \ps(d)_A$ and
it follows that $\sum_{j=1}^n \o_j^2 \geq \sum_{j=1}^n \b_j^2$, where
$(\b_j)$ is the vector of Schmidt coefficients of $\ket{\ps(n)}$ in decreasing
order.  This last sum is easy to evaluate, however:
\begin{eqnarray}\label{eq:bound}
\sum_{j=1}^n \b_j^2
&\geq& \sum_{j=1}^{\lfloor n/m \rfloor} \sum_{i=1}^m \frac{1}{jC(n)m} 
~ ~ \geq ~ ~ 1 - \frac{\log(m)}{\log(n)}.
\end{eqnarray}
Thus, for any fidelity $1-\e < 1$, the requirement $n>m^{(1/\e)}$ on $\ket{\m(n)}$
suffices.  If we view the state $\ket{\ph_{AB}}$ as a string of $\log m$ 
pairs of qubits then $\ket{\m(n)}$ need only consist of 
$(\smfrac{1}{\e}) \log m$ pairs of qubits, which is only linear in the 
number of qubits of $\ket{\ph_{AB}}$.

The embezzlement protocol we present here requires absolutely no
communication and uses the same set of catalysts for every input.
Is it possible that by tailoring the catalyst to the target state
as well as making use of local operations and classical
communication that we could find more effective embezzlement
schemes?  Not significantly.  Let $n$ be the Schmidt
rank of the catalyst $\ket{\xi}$ and consider the transformation
$\ket{\xi} \mapsto \ket{\xi} \ox \ket{\ph_{AB}}$.
Suppose the optimal LOCC protocol 
yields the state $\s_{AB}$.  In Ref.~\cite{VidalJN00} it was
shown that this optimal $\s_{AB}$ will be a pure state with Schmidt basis matching that
of $\ket{\xi}\ox\ket{\ph_{AB}}$.  Since the entanglement cannot be increased
by an LOCC protocol, $S(\s_A) \leq S(\xi_A)$.  Therefore, if 
$\Tr|\s_A - \xi_A \ox \ph_A| = \d$, the Fannes' inequality \cite{Fannes73}
implies that, for $\d < 1/\mathrm{e}$,
\begin{eqnarray}
S(\ph_A) 
&\leq& | S(\xi_A \ox \ph_A) - S(\s_A) | 
~  ~ <~ ~\d( \log(m) + \log(n) ) + \eta(\d),
\end{eqnarray}
where $\eta(\d) = -\d \log \d$ and $m$ is the rank of $\ph_A$, and hence
\begin{eqnarray} \label{eqn:Fannes}
\frac{S(\ph_A) - \eta(\d)}{\log(m)+\log(n)} &<& \d.
\end{eqnarray}
For our protocol, however, a straightforward calculation reveals 
that
\begin{eqnarray}
\d ~ ~ = ~ ~ \Tr|\o(n)_A - \m(n)_A| & = & 
\sum_{j=1}^n{(\mu^2_j-\o_j^2)} + \sum_{j=n+1}^{nm}{\o_j^2}
 ~ ~ \leq ~ ~ \frac{2\log(m)}{\log(n)},
\end{eqnarray}
where we used the fact that $\m_j \geq \o_j$ for $1\leq j \leq n$
and $\m_j=0$ for $j>n$ combined with the bound of Eq.~(\ref{eq:bound}).
Clearly, for a fixed $\ph_{AB}$ this $\d$ saturates Eq.~(\ref{eqn:Fannes}) 
to within a constant factor for large $n$.

We have shown that it possible to embezzle entanglement without any
communication whatsoever and that the set $\{\ket{\m(n)}\}$ can be used to
embezzle any bipartite pure state.  Furthermore, we have shown that the
universal family $\{\ket{\m(n)}\}$ is nearly optimal, almost saturating
the limit on embezzlement imposed by the continuity of the von Neumann
entropy.

The embezzlement phenomenon has a number of consequences for the study
of quantum information.  For example, it implies that the 
\emph{trumping relation} on bipartite entangled states \cite{DaftuarK01}
is not stable to arbitrarily small perturbations.  In other words, 
in the presence
of unrestricted catalysts, all states are effectively reachable from all
others without communication.  Similarly, a standard proof technique in
quantum communication complexity reduces distributed function evaluations
to related state transformations \cite{ClevevNT99}.  The amount of 
communication for the distributed problem is related to the 
amount of communication required to perform the corresponding state 
transformation.  Our results imply that 
this technique will fail on attempts to study the 
probabilistic communication complexity of functions when
an unlimited amount of initial entanglement is allowed.

Acknowledgments: We would like to thank Sumit Dixit for 
helpful conversations.  This work was supported in part by the 
National Science Foundation under grant no.~EIA-0086038, 
an HP/MSRI postdoctoral fellowship, the Defense 
Advanced Research Projects Agency (DARPA) and Air Force Laboratory, Air 
Force Materiel Command, USAF, under agreement number F30602-01-2-0524.

\end{document}